\documentclass[pamm,a4paper,fleqn%
]{w-art}
\usepackage{times,cite,w-thm}
\theoremstyle{plain}

\theoremstyle{definition}

\usepackage[]{graphicx}
\begin{document}
\DOIsuffix{theDOIsuffix}
\Volume{XX}
\Month{01}
\Year{2007}
\pagespan{1}{}



\title[Clustering of point vortices]{Clustering of point vortices in a periodic box}


\author[F.Author]{Makoto Umeki\inst{1,}%
  \footnote{Corresponding author\quad E-mail:~\textsf{umeki@phys.s.u-tokyo.ac.jp},
            Phone: +81\,3\,6413\,0831,
            Fax: +81\,3\,6413\,0873}}
\address[\inst{1}]{
Department of Physics, Graduate School of Science,
University of Tokyo, 7-3-1 Hongo, Bunkyo-ku, Tokyo 113-0033}
\begin{abstract}
  The Monte Carlo simulation of $N$ point vortices with square periodic boundary conditions 
  is performed where $N$ is order of 100. The clustering property is examined by 
  computing the $L$ function familiar in the field of spatial ecology. 
  The case of a positive value of $L$ corresponds to the state of 
  clustering and the Onsager's negative temperature. 
\end{abstract}
\maketitle                   





\section{Monte Carlo simulation of point vortices in a periodic box}

The motion of point vortices (PVs) in a periodic box has been studied numerically 
in \cite{Umeki1,Umeki2}. 
The $K$ and $L$ functions, which are familiar in the field of spatial ecology, 
are introduced in order to quantitate clustering of PVs. 
Periodic boundary conditions guarantee spatial homogeneity and 
we need not to include the edge corrections in the $K$ function.

In this report, the Monte Carlo simulation is performed for 100 identical PVs 
in order to obtain the probability distribution function (PDF) for the hamiltonian. 
Yatsuyanagi {\it et al.} \cite{msano2} showed the PDF for two types of 
positive and negative PVs and the clustered state of the negative temperature. 
Figure 1 shows the histogram of the hamiltonian for $N$ identical PVs
\begin{equation}
H=\sum_{i=1}^N\sum_{j=i+1}^N \{[-{\rm Re} \ln \sigma (z_i-z_j)] + \Omega |z_i-z_j|^2/2\},
\label{eq1}
\end{equation}
where $z_i$ is the complex position of the PV, ${\rm Re}$ denotes the real part, $\sigma$ is the Weierstrass sigma function 
parametrized by two numbers $\omega_1=1/2$ and $\omega_2=i/2$, 
$\Omega=\pi$, $N=100$, and the ensemble number is $10^4$. 
The $\Omega$ term denotes a rigid rotation centered at each PV. 
Similarly to \cite{msano2}, 
the distribution with decaying tails is obtained. 
The figures 2-4 and 5-7 show the spatial distribution of PVs and 
the $L$ functions for the minimum, median and maximum of the hamiltonian, respectively.  
The $L$ and $K$ functions are defined by 
\begin{equation}
L(r) = \sqrt{K(r)/\pi} -r,
\label{eq2}
\end{equation}
\begin{equation}
K(r) = (\lambda N)^{-1}\sum_{i=1}^N\sum_{j=1, \ne i}^N \theta(r-|z_i-z_j|),
\label{eq3}
\end{equation}
where $\lambda=N/S$, $S$ is the area, and $\theta$ is the step function.
For complete spatial randomness (CSR), we have $K=\pi r^2$ and $L=0$. 
The positive (negative) values of $L$ imply clustering (uniform spacing). 

The right-side tail in Figure 1 corresponds 
to clustering of PVs, the positive value of $L$, and the state of 
the Onsager's negative temperature.
The left-side tail in Figure 1 shows uniform spacing of PVs, the negative 
value of $L$, and the positive temperature.
Figure 6 shows an oscillatory behavior of the $L$ function for the median of $H$, 
where $H=2.61$ is close to the peak of the histogram ($H=2.613$). 
Since there is a minimum in the hamiltonian for each pair of $(i,j)$, 
the total hamiltonian $H$ is also lower bounded. 
However, we have no upper bound for $H$ since $\sigma(z) \sim z$ for $z\sim 0$. 
In general, it is difficult to judge whether the point distribution is 
clustered, completely spatially random, or uniformly spaced. 
The $L$ function is found to be useful for the present purpose. 

The numerical simulation of 100 PVs with various conditions has been done in 
\cite{Umeki1,Umeki2}. One of the typical behaviors observed in the case of positive
and negative PVs with the same strength is the scattering and recoupling of 
the linearly moving pair of PVs. 
A gradient method for detecting stable stationary 
configurations of PVs, which corresponds to the system of point sources
with a uniform sink, has been found by the author and will be reported elsewhere. 

\begin{figure}[ht]
\begin{center}
\includegraphics[height=55mm]{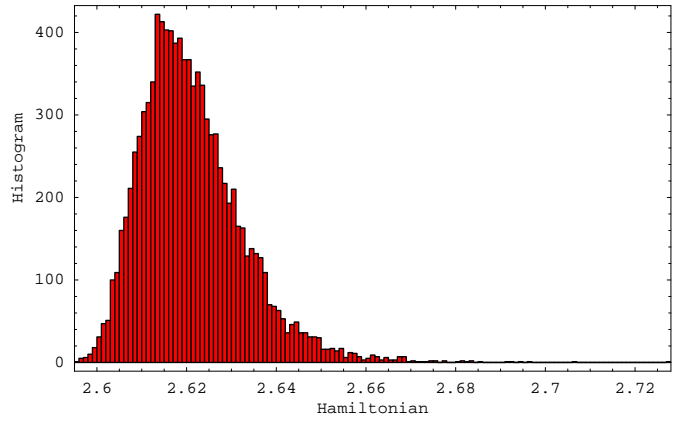}
\caption{Histogram of the hamiltonian of periodic point vortices with the Monte Carlo method.
The right(left)-side tail denotes the negative (positive) temperature state.}
\label{fig:1}
\end{center}
\end{figure}

\begin{figure}[ht]
\begin{minipage}{55mm}
\includegraphics[height=55mm]{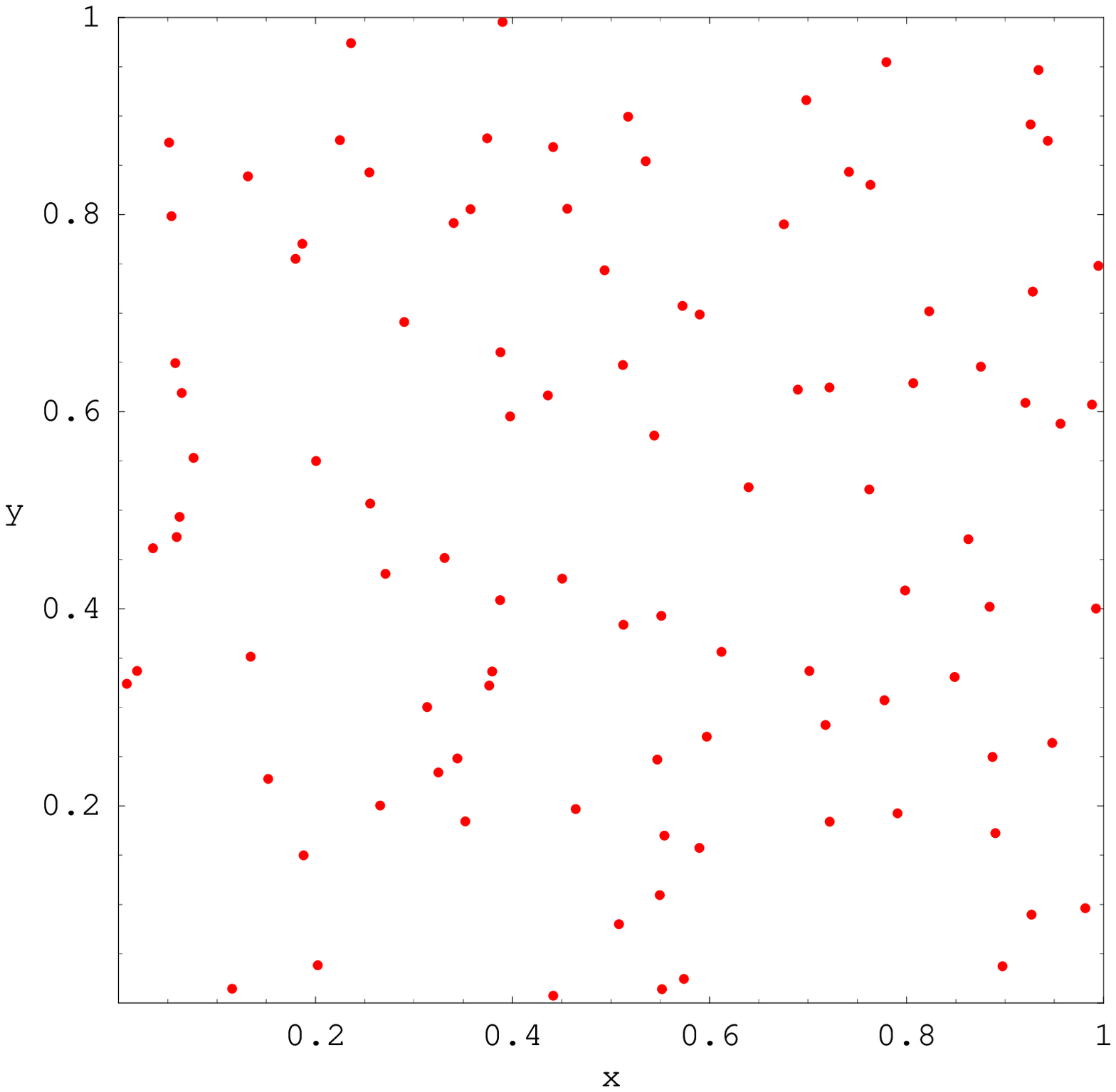}
\caption{The distribution of PVs for the minimum of the hamiltonian.}
\label{fig:2}
\end{minipage}
\hfil
\begin{minipage}{55mm}
\includegraphics[height=55mm]{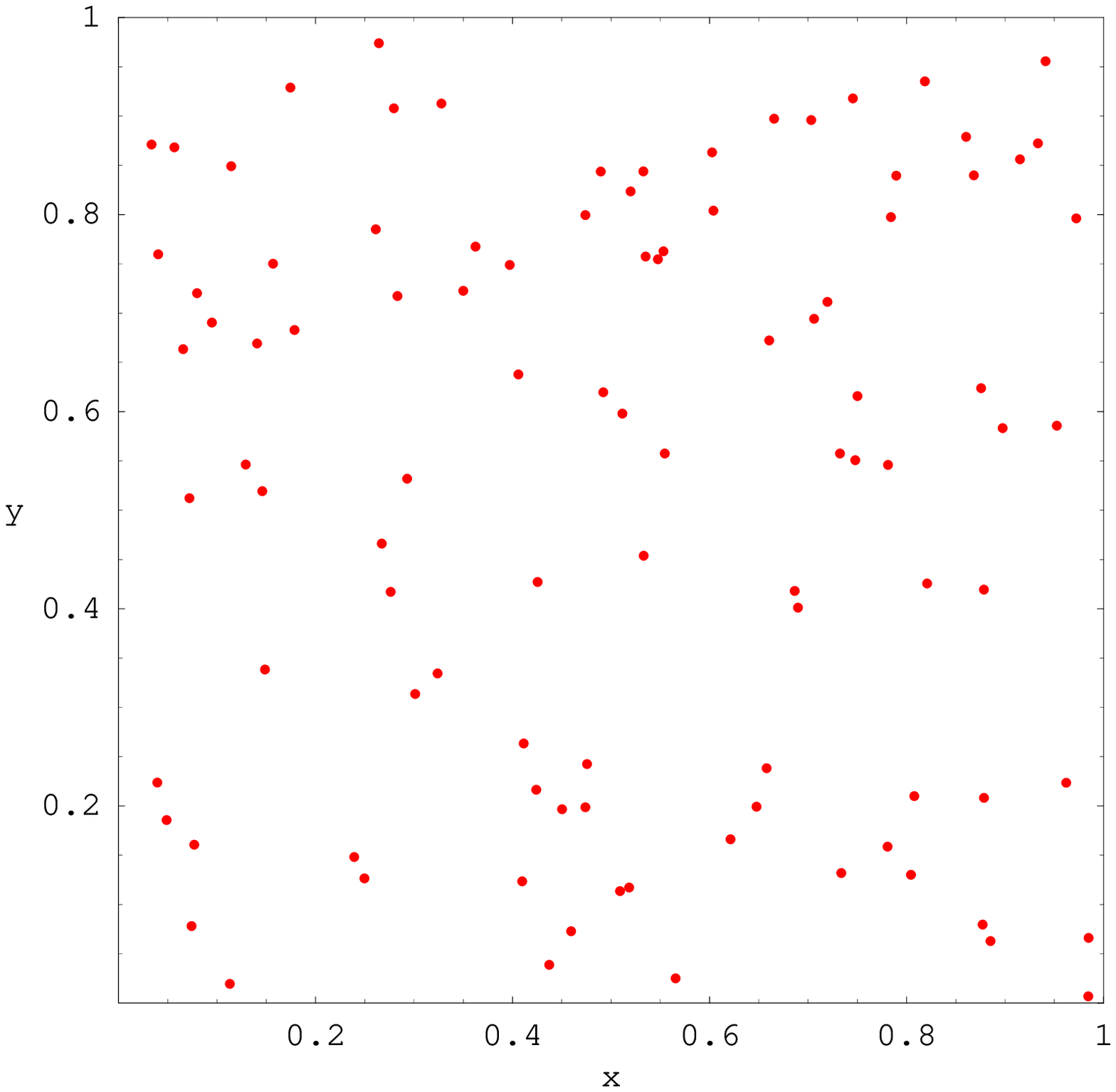}
\caption{The distribution of PVs for the median of the hamiltonian.}
\label{fig:3}
\end{minipage}
\hfil
\begin{minipage}{55mm}
\includegraphics[height=55mm]{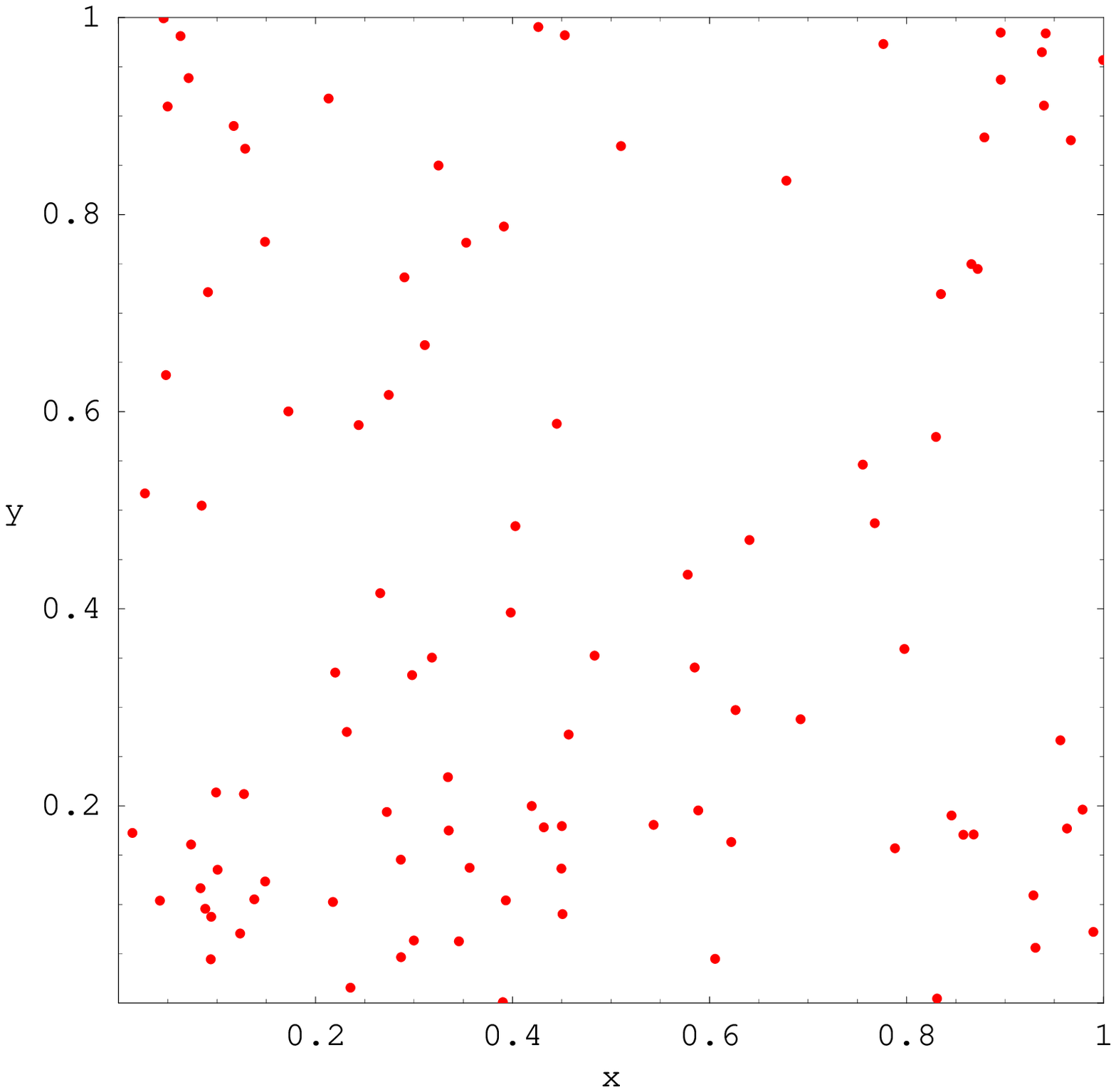}
\caption{The distribution of PVs for the maximum of the hamiltonian.}
\label{fig:4}
\end{minipage}
\end{figure}

\begin{figure}[ht]
\begin{minipage}{55mm}
\includegraphics[height=55mm]{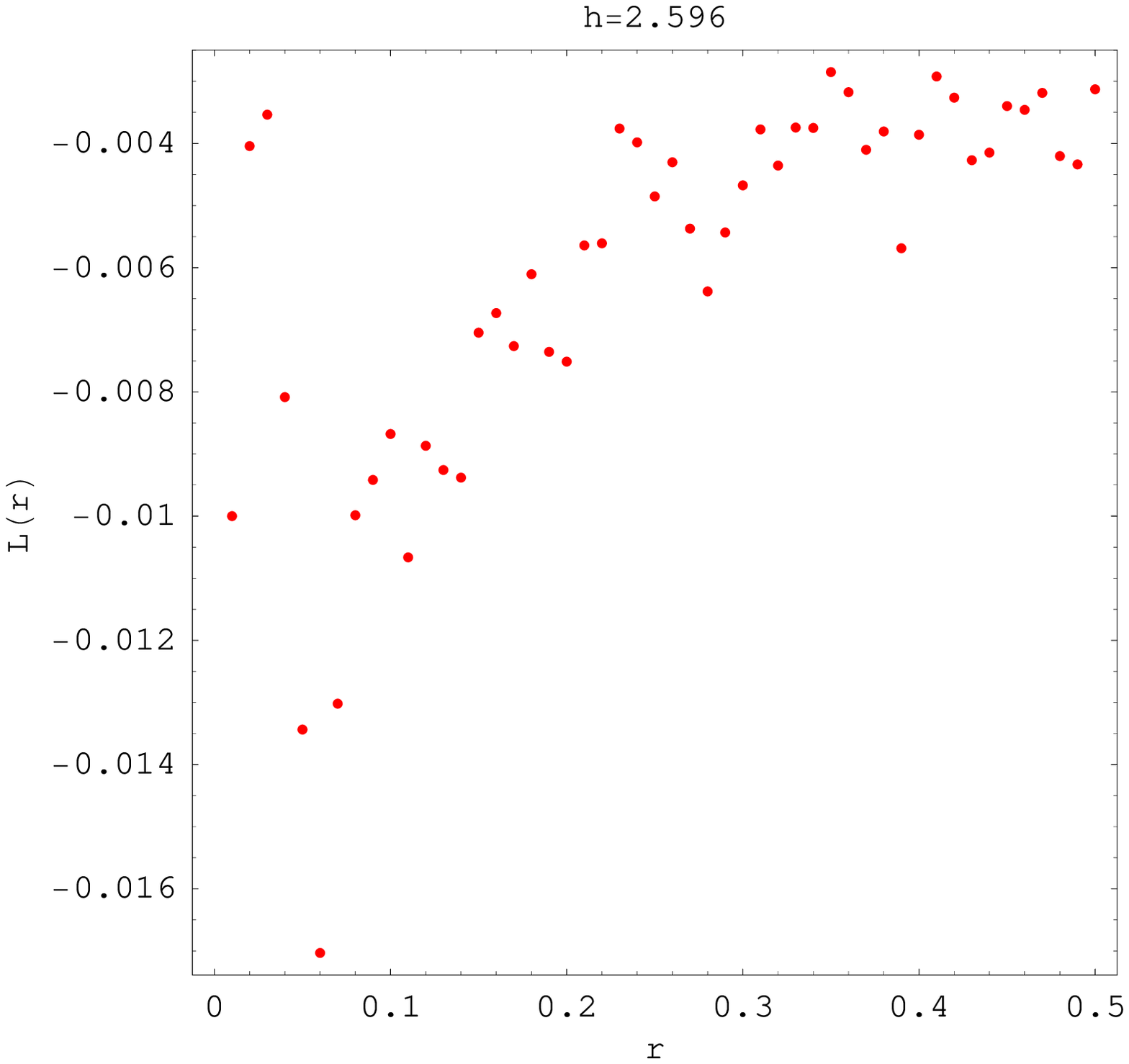}
\caption{The $L$ function for the minimum of the hamiltonian.}
\label{fig:5}
\end{minipage}
\hfil
\begin{minipage}{55mm}
\includegraphics[height=55mm]{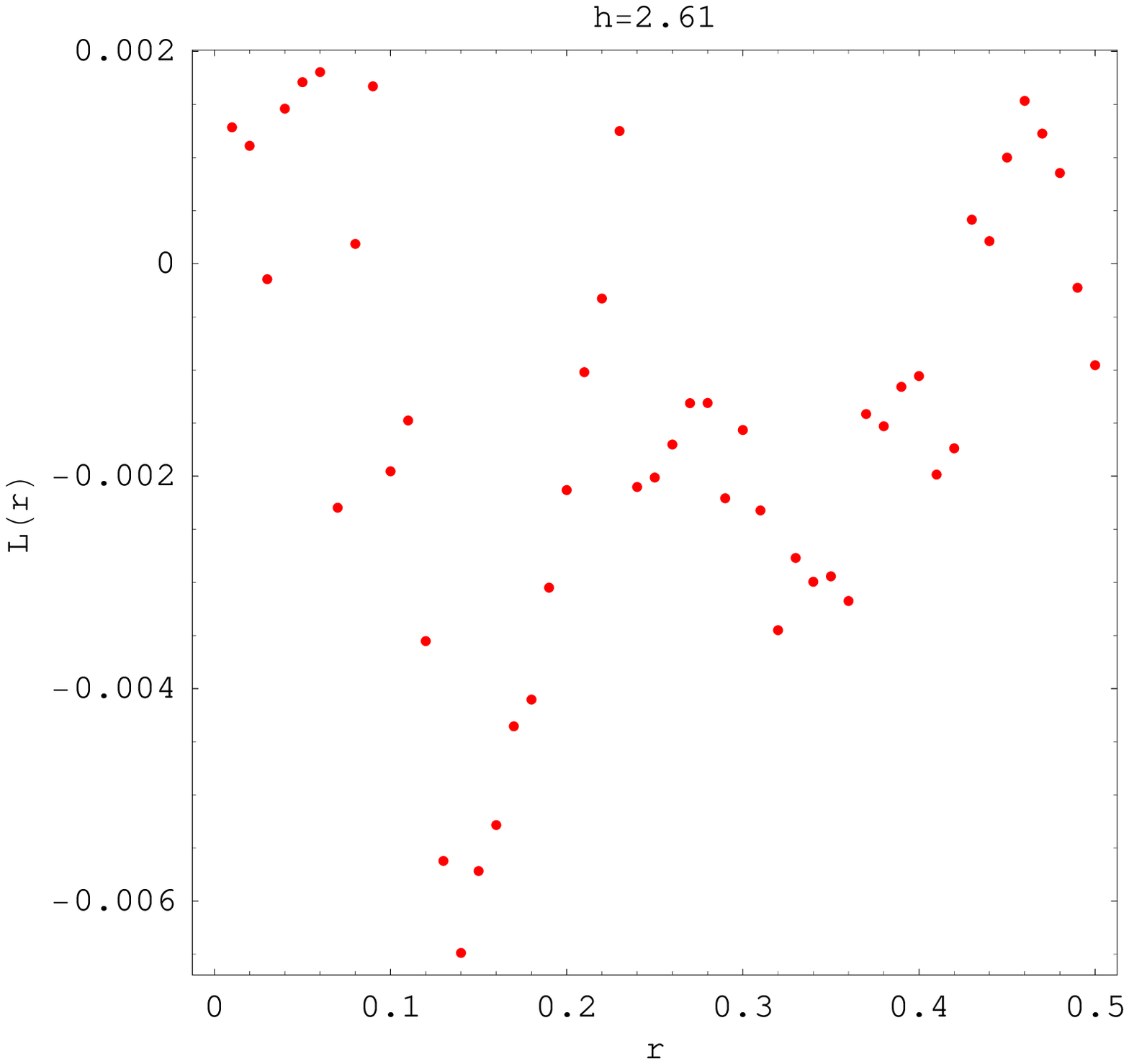}
\caption{The $L$ function for the median of the hamiltonian.}
\label{fig:6}
\end{minipage}
\hfil
\begin{minipage}{55mm}
\includegraphics[height=55mm]{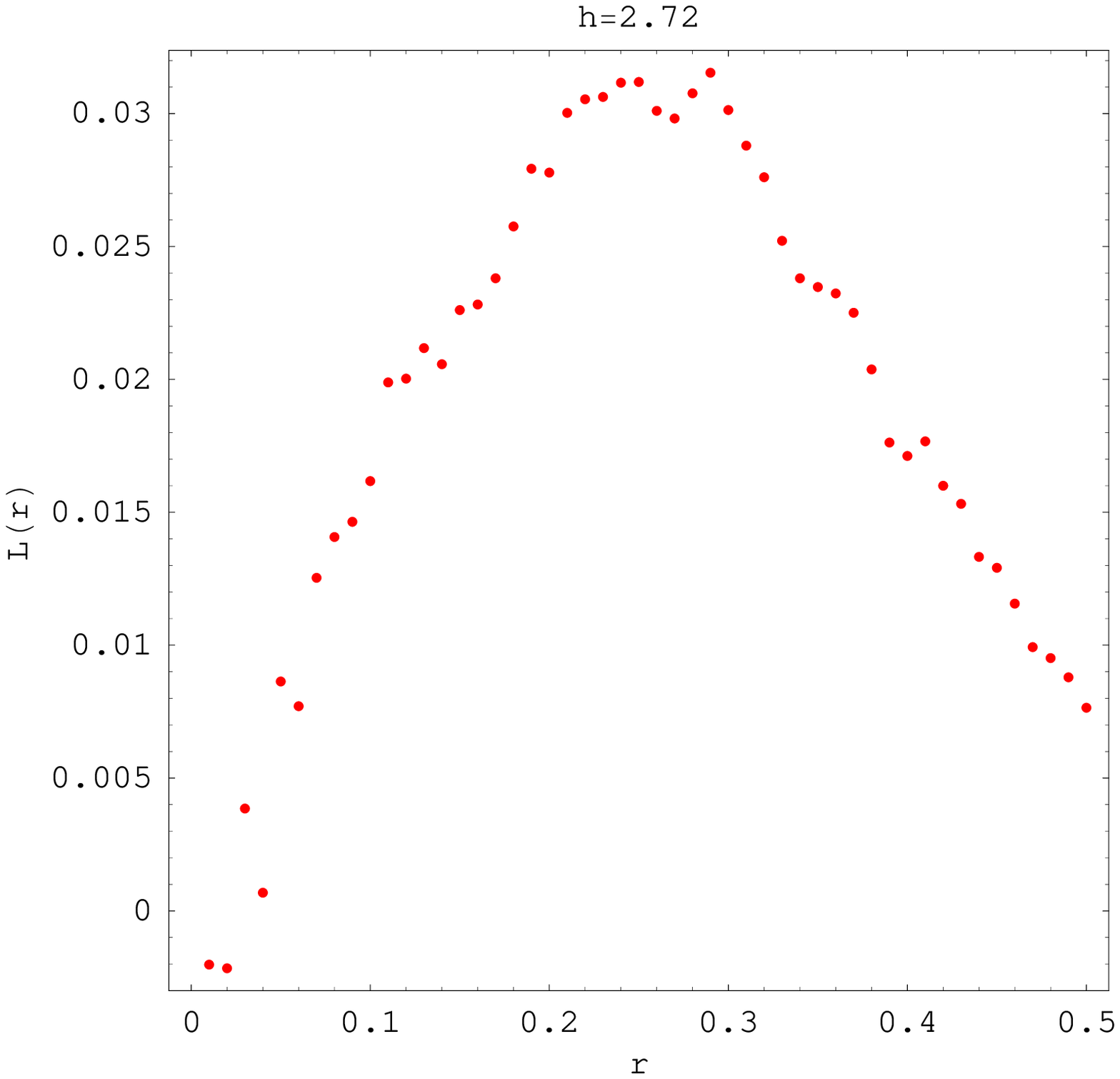}
\caption{The $L$ function for the maximum of the hamiltonian.}
\label{fig:7}
\end{minipage}
\end{figure}

\end{document}